\documentclass{elsart}

\usepackage{graphicx}

\usepackage{amssymb}

\usepackage{color}

\journal{Commun. Nonlinear Sci. Numer. Simul.}

 \newcommand{\Cont}[0] {\mbox{C}}
 \newcommand{\dtot}[2] { \frac{d {#1} } {d {#2}} }
 
 \newcommand{\dpar}[2] { \frac{\partial {#1} } {\partial {#2}} }
 \newcommand{\ndpar}[3] { \frac{\partial^{#3} {#1} } {\partial #2 ^{#3} }}
 \newcommand{\ndparh}[3] {{\partial^{#3} {#1} }/{\partial #2 ^{#3} }}

 \newcommand{\OL}[1] {\textrm{\emph{\large O}}\!\left({#1}\right)}

 \newcommand{\eqref}[1] {(\ref{#1})}

\begin{document}

\begin{frontmatter}

 \title{Numerical interactions between compactons and kovatons of the Rosenau-Pikovsky
       $K(\cos)$ equation}

\author{Julio Garral\'on},
\ead{jgarralon@uma.es}
\author{Francisco Rus},
\ead{rusman@lcc.uma.es}
\author{Francisco R. Villatoro}
\ead{villa@lcc.uma.es}
\corauth{Corresponding author: villa@lcc.uma.es (Francisco R. Villatoro).}

\address{Departamento de Lenguajes y Ciencias de la Computaci\'on, \\
Universidad de M\'alaga, Campus de Teatinos, 29071 M\'alaga,
Spain}



\begin{abstract}
A numerical study of the nonlinear wave solutions of the
Rosenau-Pikovsky $K(\cos)$ equation is presented.
This equation supports at least two kind of solitary waves with compact support:
\emph{compactons} of varying amplitude and speed, both bounded, and \emph{kovatons} which
have the maximum compacton amplitude, but arbitrary width.
A new Pad\'e numerical method is used to simulate the propagation and, with small artificial viscosity added, the interaction between these kind of solitary waves.
Several numerically induced phenomena that appear while
propagating these compact travelling waves are discussed quantitatively, including self-similar
forward and backward wavepackets. The collisions of compactons and kovatons show new phenomena such as the inversion of compactons and the generation of pairwise ripples decomposing into small compacton-anticompacton pairs.
\end{abstract}


\end{frontmatter}

\section{Introduction}
\label{Intro}

\emph{Compactons} are compactly-supported solitary waves that appear as solutions of some
generalizations of the Korteweg--de Vries (KdV) equation
characterized by the presence of nonlinear dispersion. Compactons were introduced
for the first time by Rosenau and Hyman~\cite{RosenauHyman1993} as
solutions of the $K(m,n)$ equation, given by
\begin{equation}
 \dpar{u}{t} + \dpar{u^m}{x} + \ndpar{u^n}{x}{3} = 0,
\label{eq:Kmn}
\end{equation}
where $x$ is the spatial variable and $t$ is time, for certain values of $m$ and $n$, and generalized by Khare and Cooper~\cite{KhareCooper1993} for $n=m$. This kind of solitary waves presents some fundamental similarities with the \emph{solitons} of
the KdV equation, including the
almost elastic interaction between them, but the nonlinear dispersion in the  $K(m,n)$ equation
makes compactons quite a distinct mathematical object with its own
properties~\cite{CooperEtAl1993,CooperHyman2001}, being its compact support
the most important.

Numerical analysis has proved to be essential in the study
of equations with compactly supported solitary waves. Pseudospectral~\cite{RosenauHyman1993,ChertockLevy2001},
finite differences~\cite{IsmailTaha1998,RusVillatoro2008},
Petrov-Garlerkin finite elements~\cite{deFrutosSanzSerna1995,GarralonRusEtAl2006,RusVillatoro2010},
Pad\'e~\cite{RusVillatoro2007b,MihailaEtAl2010},
and particle methods~\cite{ChertockLevy2001} have been used with more
or less success.
However, the numerical simulation of nonlinear travelling waves presents several
numerically induced phenomena such as spurious radiation,
artificial dissipation, or errors in group velocity, and the numerical
evolution of compactons, especially at the edges of the support,
is not free of these undesired effects
\cite{deFrutosSanzSerna1995,RusVillatoro2007a}. Special attention has to be paid to the behaviour of compactons under mutual collisions, a key property to check for their robustness~\cite{RusVillatoro2007a,RusVillatoro2009,CardenasEtAl2011}. Although there is a large number of nonlinear partial differential equations presenting compactly-supported solitary wave solutions~\cite{RusVillatoro2009b}, only a few have been numerically studied in detail.

Pikovsky and Rosenau~\cite{RosenauPikovsky2005} introduced a generalization of the $K(n,n)$ equation for the study of the quasicontinuous approximation of a one-dimensional lattice of coupled limit-cycle oscillators. Let us summarize the derivation presented in Refs.~\cite{PikovskyRosenau2006,PikovskyBook2001}. The $k$-th self-sustained oscillator with frequency $\omega$ is characterized by a phase $\phi_k$ that obeys $d{\phi}_k/d\tau=\omega$, where $\tau$ is time. Under weakly coupling, the equations for the phase differences read
\[
 \dtot{{\phi}_k}{\tau} = \omega + q(\phi_{k-1}-\phi_{k}) + q(\phi_{k+1}-\phi_{k}),
\]
where $q$ is the coupling function, even and $2\pi$-periodic in its argument. Introducing the variable $u_k=\phi_{k+1}-\phi_{k}$, the phase equations can be written as
\begin{equation}
 \dtot{u_k}{\tau}  = q(u_{k+1}) - q(u_{k-1}),
 \label{eq:Discrete}
\end{equation}
whose quasicontinuous approximation yields
\[
 \dpar{u}{\tau} = 2\,h\,\dpar{(q(u))}{y} + 2\,\frac{h^3}{3!}\,\ndpar{(q(u))}{y}{3}
 + \OL{h^5},
\]
where $h$ is the spatial step of the lattice. Neglecting higher order terms and applying a change of variables, this equation can be non-dimensionalized to
\begin{equation}
 \dpar{u}{t} = \dpar{q(u)}{x} + \ndpar{q(u)}{x}{3}.
\label{eq:Kfunc}
\end{equation}
For $q(u)=-\cos(u)$ we obtain the so-called $K(\cos)$ equation, given by
\begin{equation}
 \dpar{u}{t} + \dpar{\cos(u)}{x} + \ndpar{\cos(u)}{x}{3} = 0,
\label{eq:Kcos}
\end{equation}
that admits compactly-supported travelling waves such as com\-pac\-tons, kinks, and what Pikovsky and Rosenau referred to as \emph{kovatons}, which result from gluing together
a kink-antikink pair with a plateau of arbitrary width.

The closed form expression for the shape of compactons and kovatons of the $K(\cos)$ equation is not known in closed form~\cite{PikovskyRosenau2006,AhnertPikovsky2008}. Its determination by means of numerical methods requires the solution of a nonlinear eigenvalue problem~\cite{Balmforth1995}. Recently, Garral\'on and Villatoro \cite{GarralonVillatoro2012a} proposed a
procedure for the numerical quadrature of the solitary waves of the $K(\cos)$ equation, which can
be used as initial conditions for compactons and kovatons numerical evolution. These solutions are only characterized by one parameter, the velocity $c$ of the solitary wave. The nonlinear relation between the amplitude and velocity of compactons must be numerically determined.

\begin{figure}
 \begin{center}
  \includegraphics{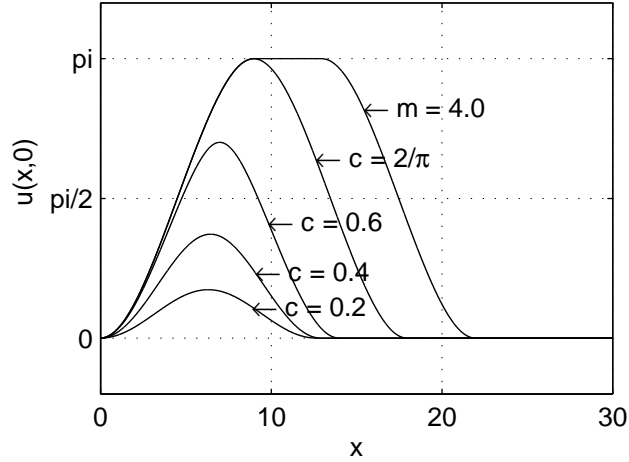}
\end{center}
  \caption{Compactons of the $K(\cos)$ equation with speeds
   $c = 0.2$, 0.4, 0.6, and $2/\pi$, and
   a kovaton with plateau length $m = 4.0$.}
  \label{fig:kcosVelos}
 \end{figure}

The behaviour of compactons (and anticompactons) of the $K(\cos)$ equation differs from those of the $K(2,2)$ equation in several aspects~\cite{GarralonVillatoro2012a}. Firstly, the absolute value of their velocity is upper bounded by $c_{\max}=2/\pi$, the velocity of the kovatons (and antikovatons), whereas for the $K(2,2)$ there is no upper bound. Secondly, compactons (anticompactons) propagates to the left (right), i.e., to minus (plus) infinity in $x$, and the opposite occurs for those of the $K(2,2)$. And thirdly, the width of the $K(\cos)$ compactons increases as it does the absolute value of the velocity, as shown in Fig.~\ref{fig:kcosVelos}, while it is independent of the velocity for the $K(2,2)$ ones.

The $K(\cos)$ equation has several applications in physical problems such an
array of superconducting Josephson junctions~\cite{LikhaevBook1991} and the
spontaneous otoacoustic emissions in the inner ear of certain
lizards~\cite{VilfanDuke2008}. Let us also note that compact kinks
are the solutions of nonlinear evolution equations modelling several
applications such as waves in viscoelastic
solids~\cite{DestradeEtAl2009}, double-stranded DNA
molecules~\cite{SaccomandiSgura2006}, and nonlinear electrical
transmission lines~\cite{ComteMarquie2006}, to mention only a few.

The contents of this paper are as follows. Next section presents a new numerical
scheme used to solve the $K(\cos)$ equation, based on Pad\'e approximants in
space and a method of lines in time. Both compactons and kovatons have
been studied, analyzing the residual structures that appear when propagating
them alone as well as after pairwise interactions in Section~\ref{Results}.
Finally, the last section is devoted to some conclusions.

\section{Numerical method}
\label{method}

Let us consider a method of lines in time and
a Pad\'e approximation method in space with periodic boundary
conditions~\cite{MihailaEtAl2010} for solving Eq.~\eqref{eq:Kcos} numerically.
Taking a fixed grid spacing $\Delta x$, the $M+1$ spatial
nodes of the interval $x\in[0,L]$ are $x_j=j\,\Delta x$, for $j=0,
1 \ldots M$, and the Pad\'e method is written as
\begin{equation}
 \label{eq:Pade}
 {\mathcal{A}(\mbox{E})}\,\dtot{U_{j}}{t}
         + c_0 \, {\mathcal{B}(\mbox{E})} \, U_j
         + \left({\mathcal{B}(\mbox{E})}
         + {\mathcal{C}(\mbox{E})}\right) \,\cos(U_j)
          = 0,
\end{equation}
where $U_j(t)\approx u(x_j,t)$, $\mbox{E}$ is the shift operator, i.e.,
$\mbox{E}\,U_j=U_{j+1}$, and the first and third derivatives are
rationally approximated by means of
$\mathcal{B}(\mbox{E})/\mathcal{A}(\mbox{E})$ and
$\mathcal{C}(\mbox{E})/\mathcal{A}(\mbox{E})$, respectively, where

\[
 {\mathcal{A}(\mbox{E})}
 =
  \frac{ \mbox{E}^{-2}+26\,\mbox{E}^{-1}+66+26\,\mbox{E}^{1}+\mbox{E}^{2}
       }{120},
\]
\[
 {\mathcal{B}(\mbox{E})}
 =
 \frac{ -\mbox{E}^{-2}-10\,\mbox{E}^{-1}+10\,\mbox{E}^{1}+\mbox{E}^{2}
      }{24\,\Delta x},
\]
\[
 {\mathcal{C}(\mbox{E})}
 =
 \frac{ -\mbox{E}^{-2}+2\,\mbox{E}^{-1}-2\,\mbox{E}^{1}+\mbox{E}^{2}
      }{2\,\Delta x^3}.
\]
The numerical method~\eqref{eq:Pade}
is fourth-order accurate in space for regular
enough solutions ($u(x,t)\in\Cont^7$), since its truncation error
terms are given by
\[
  \mbox{TET}\{u(x,t)\}
  = \frac{\Delta x^4}{240} \,
 \ndpar{}{x}{7}\cos(u)
 +\OL{\Delta x^6}.
\]

It should be noted that the Pad\'e numerical method~\eqref{eq:Pade} is a fourth-order approximation in $\Delta x$ to the continuous equation~\eqref{eq:Kcos}, which in turn is a fourth-order approximation in $h$ to the discrete equation~\eqref{eq:Discrete}. Although, $\Delta x$ and $h$ can be related by a proper scaling, physically it has no sense, since $h$ is a fixed parameter of the original discrete model, but $\Delta x$ is a tunable numerical parameter.

In solutions of the $K(\cos)$ equation with multiple colliding
compactons and kovatons, ripples (or nonsmooth solutions) are developed
reducing the effective order of accuracy and introducing numerical
instabilities which may blow up the
solution~\cite{deFrutosSanzSerna1995,RusVillatoro2007b,RusVillatoro2007a}.
In order to avoid these instabilities,
an artificial viscosity term, $\mu\,\ndparh{u}{x}{4}$,
with $\mu$ small enough, must be introduced
into the non-dissipative method given by Eq.~\eqref{eq:Pade} and
numerically discretized by means of a second-order accurate
five-point difference formula, given by
\begin{equation}
 \label{DEequation}
  \mathcal{D}(\mbox{E})\,U_j
   =
  \frac{\mbox{E}^{-2}-4\,\mbox{E}^{-1}+6-4\,\mbox{E}^{1}+\mbox{E}^{2}
       }{\Delta x^4}\,U_j.
\end{equation}
Note that the addition of the artificial viscosity reduces the order of consistency of the method down to the second, resulting in
\[
  \mbox{TET}_{\mathcal{D}}\{u(x,t)\}
  = \frac{\Delta x^2}{6} \,
 \ndpar{u}{x}{6}
 +\OL{\Delta x^4}.
\]

The second-order implicit midpoint rule is used for the
integration in time, yielding

\begin{eqnarray}
 &&
 {\mathcal{A}(\mbox{E})}\,
    \frac{U_j^{m+1}-U_j^{m}}{\Delta t}
         +( \mu\,\mathcal{D}(\mbox{E}) + c_0\, {\mathcal{B}(\mbox{E})}
              )\,
          \left(\frac{U_j^{m+1}+U_j^{m}}{2}\right)
 \nonumber \\ &&
    \phantom{  {\mathcal{A}(\mbox{E})}\,
    \frac{U_j^{m+1}-U_j^{m}}{\Delta t} }
         + ({\mathcal{B}(\mbox{E})}+ {\mathcal{C}(\mbox{E})}) \,
         \cos\,\left(\frac{U_j^{m+1}+U_j^{m}}{2}\right)
          = 0,
 \label{Midpoint}
\end{eqnarray}

where $U_j^m$ is an approximation to $u(x_j,t^m)$, with
$t^m=m\,\Delta t$.

\section{Presentation of results}
\label{Results}

Extensive numerical experiments with the above scheme has proved to be good in accuracy and conservation properties for both $\mu=0$ and, when required, a small properly chosen $\mu$. Let us summarize the results in the following subsections.

\subsection{One-compacton solution}
\label{ssec:OneCompacton}

Let us consider the numerical simulation of the propagation of one compacton of the $K(\cos)$ equation
calculated by method~\eqref{Midpoint} without artificial viscosity ($\mu=0$), stopped in the integration interval by setting $c_0=c$.

\begin{table}
\[
 \begin{array}{c|c|c|c|c|c}
 \cline{1-6}
   \multicolumn{1}{c|}{} & \multicolumn{5}{c}{\Delta x, \quad \Delta t = 0.001}   \\
  \cline{2-6}
  c & 0.1 & 0.05 & 0.01 & 0.005 & 0.001   \\
 \hline
    1/2 & 7.9\times 10^{-6}    & 3.5\times 10^{-6}  & 6.1\times 10^{-7}  & 1.2\times 10^{-6}  & - \\
  2/\pi &  1.6\times 10^{-5}   &  8.0\times 10^{-6} & 3.0\times 10^{-6} & 4.5\times 10^{-6} & - \\
 \hline
 \hline
   \multicolumn{1}{c|}{} & \multicolumn{5}{c}{\Delta x=0.1, \quad \Delta t}   \\
 \cline{2-6}
  c & 0.1 & 0.05 & 0.01 & 0.005 & 0.001  \\
 \hline
   1/2 &  7.7\times 10^{-6}  &  7.8\times 10^{-6} & 8.7\times 10^{-6} & 8.1\times 10^{-6} & 7.9\times 10^{-6}  \\
  2/\pi &  1.0\times 10^{-5}  &  1.4\times 10^{-5} & 1.6\times 10^{-5} & 1.6\times 10^{-5} & 1.6 \times 10^{-5}\\
  \hline
  \end{array}
\]
 \caption{Numerical errors in infinite norm for the one-compacton solution
 of the $K(\cos)$ equation at $t = 20$  as a function of $\Delta x$ and fixed $\Delta t=0.001$ (top), and as a function of $\Delta t$ and fixed $\Delta x=0.1$ (bottom), using $L=150$, $\mu=0$, and $c_0=c$.}
\label{tab:ErrXT}
\end{table}

Table~\ref{tab:ErrXT} shows the error in infinite norm between the exact one-compacton
solution (determined as described in Ref.~\cite{GarralonVillatoro2012a} with machine precision) and the numerical one at $t=20$, i.e., $\max_j |U_j(20) - u(x_j,20)|$, as a function of $\Delta x$ (top) and $\Delta t$ (bottom). For $\Delta x\ge 0.01$ and $\Delta t=0.001$, Table~\ref{tab:ErrXT} shows that the numerical error decreases as $\Delta x$ does for compactons with velocities $c=0.5$ and $c=2/\pi$ (the highest one). For $\Delta x< 0.01$ the error is higher for $\Delta x\le 0.001$ the solution blows up even for a stopped compacton (indicated as $-$ in the table); apparently this behaviour is surprising since the CFL (Courant-Friedrichs-Lewy) condition ($\Delta x/\Delta t \ge c-c_0 = 0$) is always met. However, the CFL condition for the numerically induced wavepackets of radiation, to be studied later in this section, does not holds, hence an instability develops. On the other hand, Table~\ref{tab:ErrXT} (bottom) shows that the error for both compactons is practically independent of $\Delta t$, and no problems of stability due to the CFL condition arise. Finally, let us note that the behaviour of the errors for anticompactons is the same as that for compactons with the same parameters (except that $c_0 = - c$ must be used to stop them in the numerical frame).

The numerical method used in this paper preserves exactly the first invariant for the $K(\cos)$ equation with $\mu=0$, i.e.,
\[
 I_1=\int^{\infty}_{-\infty} u(x,t)\,dx,
\]
since by summing in space it may be easily shown that $\sum_{m} u^n_m = \sum_{m} u^0_m$. However, second invariant
\[
 I_2=\int^{\infty}_{-\infty} \sin(u(x,t))\,dx
\]
is not exactly preserved. This invariant has been numerically calculated by means of the trapezoidal quadrature rule.

\begin{table}
\[
 \begin{array}{c|c|c|c|c|c}
 \cline{1-6}
   \multicolumn{1}{c|}{} & \multicolumn{5}{c}{\Delta x, \quad \Delta t = 0.001}   \\
  \cline{2-6}
  c & 0.1 & 0.05 & 0.01 & 0.005 & 0.001   \\
 \hline
  1/2  & 1.4\times 10^{-13}    & 2.4\times 10^{-12}  & 9.5\times 10^{-10}  & 2.9\times 10^{-9} & -   \\
 2/\pi &  2.9\times 10^{-13}   &  1.5\times 10^{-11} & 1.2\times 10^{-9} & 2.6\times 10^{-9} & -   \\
 \hline
 \hline
   \multicolumn{1}{c|}{} & \multicolumn{5}{c}{\Delta x=0.1, \quad \Delta t}   \\
 \cline{2-6}
  c & 0.1 & 0.05 & 0.01 & 0.005 & 0.001  \\
 \hline
  1/2 & 1.7\times 10^{-12}    & 1.0\times 10^{-12}  & 1.4\times 10^{-13}  & 1.5\times 10^{-12} & 4.7\times 10^{-13} \\
  2/\pi &  7.1\times 10^{-15}   &  1.1\times 10^{-12} & 2.3\times 10^{-12} & 9.7\times 10^{-13} & 2.9\times 10^{-13}\\
 \hline
  \end{array}
\]
 \caption{Numerical error for the second invariant $I_2$ for the one-compacton
 solution of the $K(\cos)$ at $t = 20$, as a function of $\Delta x$ and fixed $\Delta t=0.001$ (top), and as a function of $\Delta t$ and fixed $\Delta x=0.1$ (bottom), using $L=150$, $\mu=0$, and $c_0=c$.}
\label{tab:ErrIXT}
\end{table}

Table~\ref{tab:ErrIXT} shows the error of the second invariant for one-compacton solution at $t=20$ as a function of $\Delta x$ (top) and $\Delta t$ (bottom), i.e.,  $|I_2(20) - I_2(0)|$. As shown in this table, the error grows as $\Delta x$ decreases for compactons with both velocities $c=0.5$ and $c=2/\pi$, except in the case with the lowest $\Delta x$ for $c=2/\pi$. By contrast, Table~\ref{tab:ErrIXT} (bottom) shows that the error of the second invariant is practically constant, nearly independent of $\Delta t$.


The numerical evolution of compactons of the $K(\cos)$ equation shows the presence of two numerically-induced small amplitude wavepackets, one propagating to the left, the \emph{forward radiation}, and the other one to the right, the \emph{backward radiation} (recall that compactons travel to the left). Extensive numerical experiments show that the velocities of the wavefront for both the backward and forward radiation are independent of $\Delta x$ and $\Delta t$ parameters. On the other hand, the amplitude of radiations emitted for compactons increases as so do $\Delta x$ and $c$.

\begin{figure}
 \begin{center}
  \includegraphics[width=12.5cm]{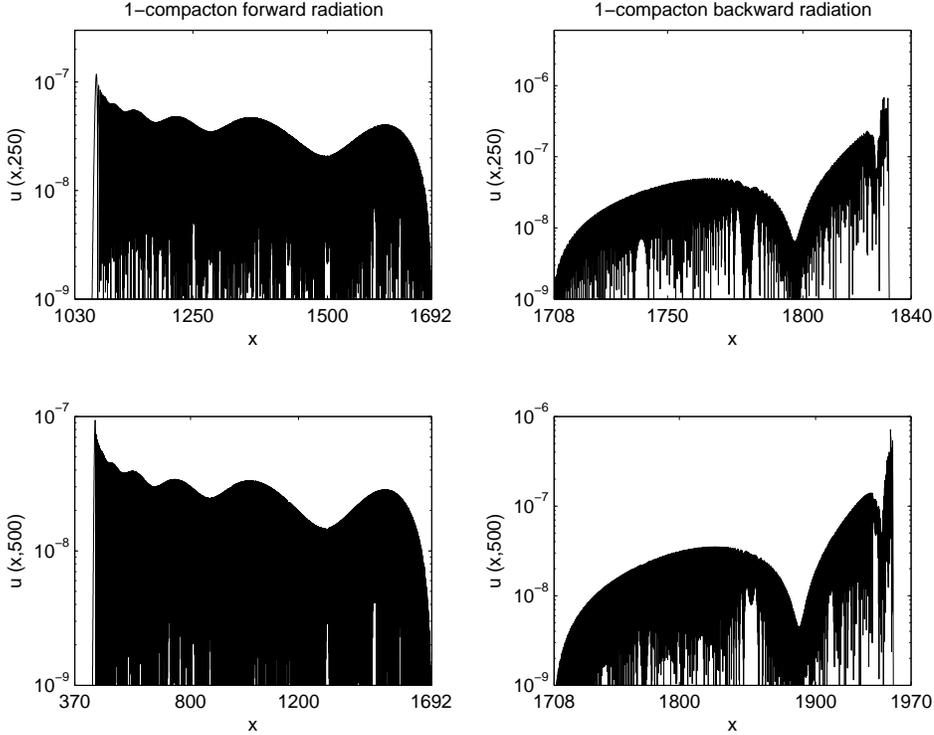}
 \end{center}
  \caption{Forward (left plots) and backward (right plots) radiations
  generated during a compacton propagation with $c=0.5$,
  $\Delta x= 0.05$, $\Delta t = 0.01$, and $\mu=0$,
  at time $t=250$ (top plots) and $t=500$ (bottom plots),
  highlighting their self-similarity.}
  \label{fig:estiradaBack}
 \end{figure}

Another worth noting fact is that the radiations shape is self-similar, as in the Rosenau-Hyman $K(2,2)$ equation case reported in Ref.~\cite{RusVillatoro2007a}. Both the forward and backward wavepackets present this self-similarity regardless of parameters $\Delta x$ and $\Delta t$. Figure~\ref{fig:estiradaBack} clearly shows this property with plots of the radiations at two different times, stretching the horizontal
axis in order to highlight the self-similarity.

Similar results have been obtained for compactons with different amplitudes in long time simulations, enlarging the computational interval in order to avoid that the wavepackets overlap due to the periodicity of the boundary conditions.

\begin{table}
{\small
\[
\begin{array}{c|c|c|c|c}
  \hline  %
   & \multicolumn{2}{c|}{c=1/2}& \multicolumn{2}{c}{c=2/\pi} \\
  \cline{2-5}
   \Delta x/ \Delta t &  c_f       & -c_b &  c_f       & -c_b      \\
\hline
   0.5  & 2.511 &  0.4938  & 3.218 & 0.6362\\
    5   & 2.511 &  0.4989  & 3.209 & 0.6360 \\
   50   & 2.511 &  0.4985  & 3.209 & 0.6362  \\
\hline
 \end{array}
\]
}
 \caption{Front speeds of both the backward ($c_b$) and forward
($c_f$) radiation relative to that of a compacton with speeds $c=1/2$ and $c=2/\pi$.
The parameters used are $\Delta x=0.05$,  $L=300$ and $c_0=c$.}
\label{tab:frentes}
\end{table}

Table~\ref{tab:frentes} shows the front speed of the forward ($c_f$) and backward ($c_b$) wavepackets of radiation relative to that of two compactons with speeds $c=0.5$ and $c=2/\pi$. These speeds have been calculated by using linear regression of the position of the wavefronts in time, calculated by using a threshold equal to the half of the maximum amplitude of the radiation in the integration interval (the same procedure used in Ref.~\cite{RusVillatoro2007a}). Table~\ref{tab:frentes} shows that both the forward ($c_f$) and backward ($c_b$) front velocities are nearly constant (independent of both $\Delta t$ and $\Delta x$) being $c_f \approx 5\,c$ and $c_b\approx -c$.

Due to the fact that the forward speed $c_f$  is five times faster than that of a compacton, care must be taken when setting parameters $\Delta x$ and $\Delta t$ in order to meet the CFL condition, so that $\Delta x/\Delta t \ge c_f \approx 5\,c$. This could be the reason of the blow up in the simulations of the last column of Tables~\ref{tab:ErrXT} and~\ref{tab:ErrIXT} (top).

\subsection{One-kovaton solution}

Let us consider the numerical simulation of the propagation of one kovaton of the $K(\cos)$ equation
calculated by method~\eqref{Midpoint} without artificial viscosity ($\mu=0$), stopped in the integration interval by setting $c_0=c$.

\begin{table}
\[
 \begin{array}{c|c|c}
 \hline
  m &  \|U(150) - u(\cdot,150)\|_\infty &    |I_2(150) - I_2(0)| \\
 \hline
 0.1 & 7.45\times 10^{-6} &  3.75\times 10^{-10} \\
 0.4 & 1.50\times 10^{-5} &  7.43\times 10^{-10} \\
 0.6 & 1.42\times 10^{-5} &  8.23\times 10^{-10} \\
 1   & 1.43\times 10^{-5} &  4.75\times 10^{-9}  \\
 4   & 1.58\times 10^{-5} &  2.81\times 10^{-9}  \\
 6   & 2.40\times 10^{-5} &  7.16\times 10^{-11} \\
 10  & 4.03\times 10^{-5} &  1.33\times 10^{-9}  \\
 40  & 1.62\times 10^{-4} &  1.06\times 10^{-9}  \\
 60  & 2.44\times 10^{-4} &  2.58\times 10^{-9}  \\
  \hline
  \end{array}
\]
 \caption{Numerical errors in infinite norm for the one-kovaton solution
 of the $K(\cos)$ equation at $t = 150$ and for the second invariant as a function of the plateau size $m$,
 using $\Delta x=0.05$, $\Delta t=0.01$, $L=220$, $\mu=0$, and $c_0=c_{\max}$.}
\label{tab:ErrM}
\end{table}

Table~\ref{tab:ErrM} shows the error in infinite norm between the exact one-kovaton
solution and the numerical one at $t=150$ as a function of the plateau size $m$. The error is nearly constant for $m<4$, but grows linearly for $m\ge 4$; in fact, a linear regression shows that, in such a case, the error is equal to $(4.1\,m-0.49)\times 10^{-6}$. Table~\ref{tab:ErrM} also shows that the error in the second invariant is very small and nearly independent of $m$.

\begin{figure}
 \begin{center}
  \includegraphics[width=12.5cm]{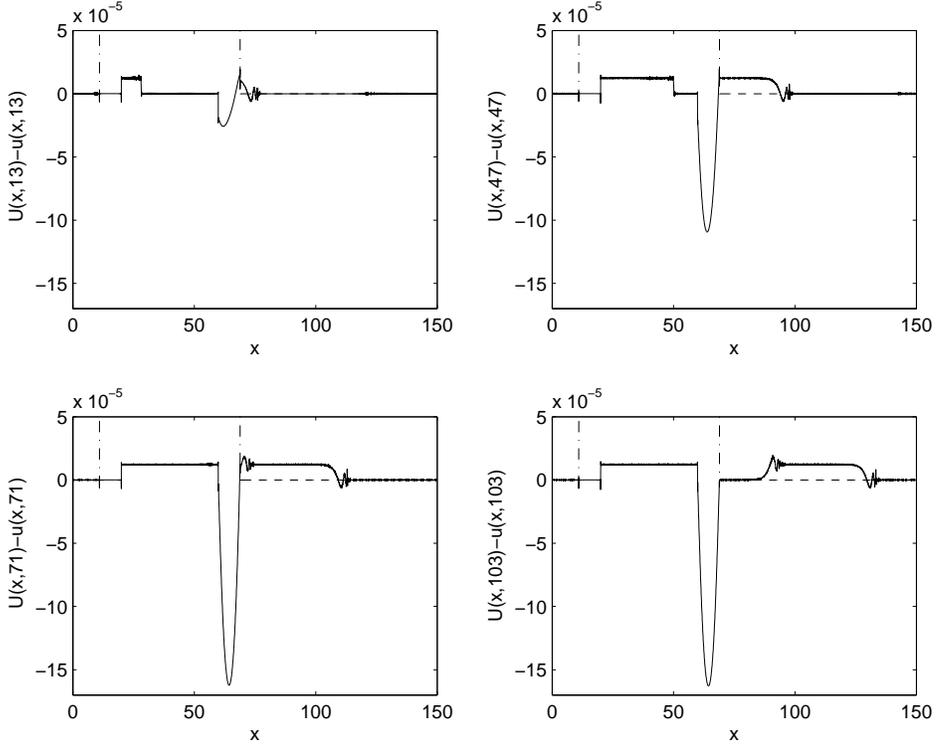}
\end{center}
  \caption{Evolution in time of the numerical error during one-kovaton propagation for $t=13$ (top left plot), $t=47$ (top right one), $t=71$ (bottom left one), and $t=103$ (bottom right one), with $m=60$, $\Delta x= 0.05$ and $\Delta t = 0.01$, $c_0=0$, and $\mu=0$.}
  \label{fig:residue}
 \end{figure}

The linear behaviour of the error for a kovaton with $m\ge 4$ shown in Table~\ref{tab:ErrM} is due to the appearance of a peak in its backward kink whose amplitude is dependent only on the value of $m$, as illustrated in Fig.~\ref{fig:residue}. The evolution of the error presents three main features occurring simultaneously with the propagation of the kovaton, whose edges correspond to the vertical, broken-and-dotted lines in Fig.~\ref{fig:residue}. First, a square pulse appears on the flat-top of the kovaton (Fig.~\ref{fig:residue}, top left plot), whose right front propagates from the forward kink to the backward one with the kovaton's speed (Fig.~\ref{fig:residue}, top right plot) until the kovaton has advanced a distance equal to $m$ (Fig.~\ref{fig:residue}, bottom left plot); here on, the shape of the square pulse remains constant without changes (Fig.~\ref{fig:residue}, bottom right plot). Second, a downward peak appears at the position of the backward kink (Fig.~\ref{fig:residue}, top left plot), whose amplitude grows until reaching a maximum value when the square pulse has passed through the flat-top (Fig.~\ref{fig:residue}, bottom left plot); from this point on, the shape of the peak remains constant. And third, another square-like pulse, but with dispersive fronts at both edges, appears behind the kovaton (Fig.~\ref{fig:residue}, top right plot) and propagates to the right, growing in width until it equals the length of the plateau, when it leaves the backward kink (Fig.~\ref{fig:residue}, bottom right plot). The conservation of the first invariant of the $K(\cos)$ equation implies that the total area of the peak is equal to the sum of the areas of both square pulses.

Kovaton propagation is also accompanied by two self-similar wavepackets of radiation whose shape is very similar to that of the one-compacton solution with maximum speed (cf. Fig.~\ref{fig:estiradaBack}). The front speed of the forward ($c_f$) and backward ($c_b$) wavepackets of radiation are nearly constant, independent of both $\Delta t$, $\Delta x$, and the width $m$ of the plateau, being approximately equal to $c_f \approx 5\,c$ and $c_b\approx -c$.

\subsection{Compacton and kovaton collisions}
\label{Results:CKcollisions}

The numerical evolution of compactons and kovatons interactions is a more
subtle problem than that of propagating them alone due to the large slopes
developed, and a stable method for one-compacton or
one-kovaton evolution may become unstable in mutual interaction.
In fact, without artificial viscosity, $\mu=0$, the numerical method blows up
during any kind of interaction between them. On the other hand,
artificial dissipation greater than zero generates tails and lags distorting
the original signal.
Numerical experimentation has proved a value of $\mu = 10^{-3}$ to be adequate.

\begin{figure}
 \begin{center}
  \includegraphics[width=12.0cm]{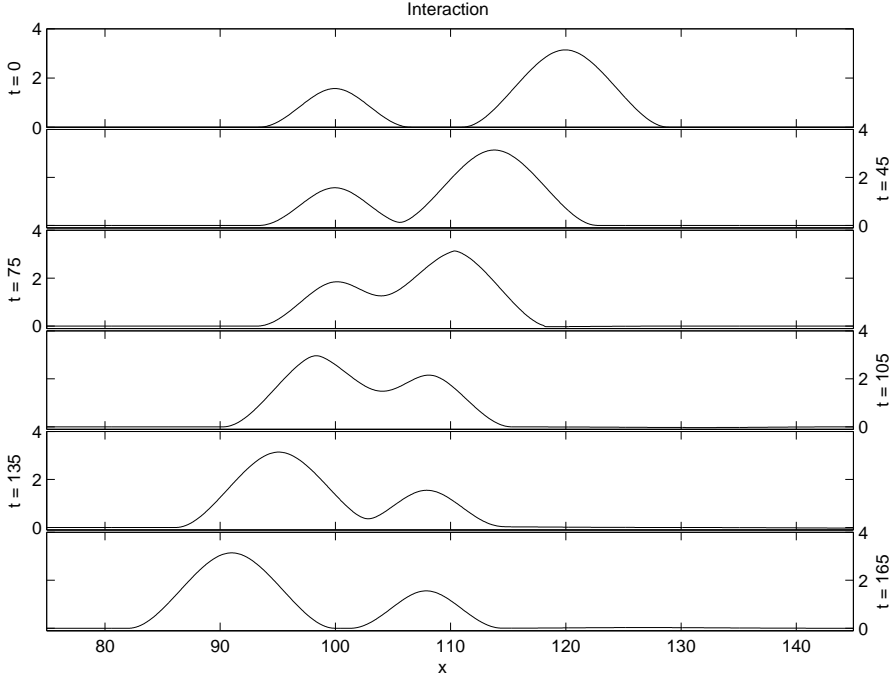}
 \end{center}
\caption{Collision between two compactons with velocities $c_1=1/2$ (the left one in the first plot)
   and $c_2=2/\pi$ (the right one), using $\mu=10^{-3}$, $c_0 =c_1$, $\Delta x = 0.05$ and $\Delta t=0.01$.}
\label{fig:CpCp}
\end{figure}

Figure~\ref{fig:CpCp} shows a sequence of plots with a collision between two compactons, one with speed $c_1=1/2$ (the left one in the first snapshot),
stopped in the computational frame by taking $c_0=c_1$, and the other one with the highest possible speed, $c_2=2/\pi$ (the right one in the same snapshot).
Both compactons collide elastically reemerging with the same shape after the interaction with a small shift in their positions,
the largest compacton suffering a positive shift to the left, and the smallest one a negative one to the right (let's compare the position of the smallest one between the top and bottom plots in Fig.~\ref{fig:CpCp}). In addition, a residual (not shown in the plot due to its small amplitude) appears at the position of the interaction.

\begin{figure}
 \begin{center}
  \includegraphics[width=12.0cm]{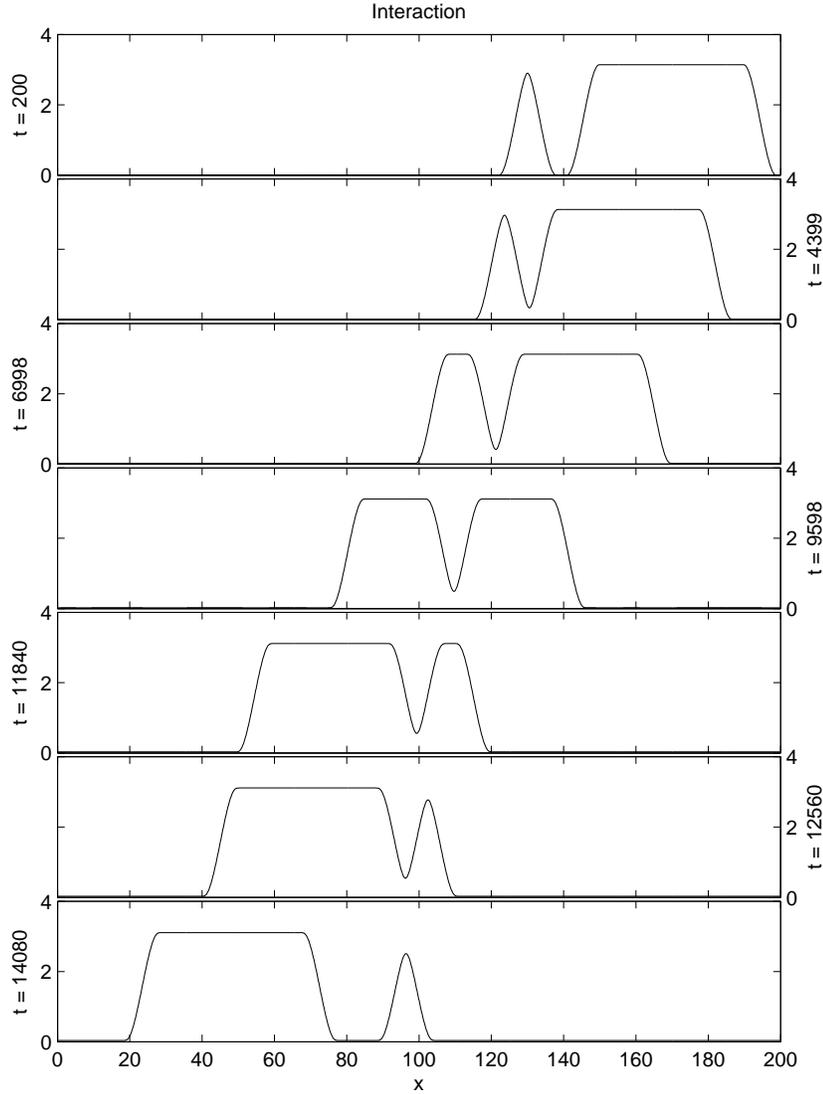}
 \end{center}
\caption{
  Collision between a compacton with speed $c=0.9 \, c_{\max} $ and a kovaton
  with plateau $m=40$, using
  $\mu=10^{-3}$, $c_0 = c$, $\Delta x = 0.05$ and $\Delta t = 0.01$.}
\label{fig:CpKv}
\end{figure}

Figure~\ref{fig:CpKv} shows a interaction between a compacton with speed $c=0.9\,c_{\max}$ and a kovaton with plateau length $m = 40$.
The compacton passes through the rising kink of the kovaton,
then it reappears reversed (second snapshot in Fig.~\ref{fig:CpKv}),
travelling like an anticompacton along the plateau (fourth snapshot),
until reaching the falling kink of the kovaton (sixth snapshot),
when it reemerges with its original shape (seventh snapshot).
Of course, this phenomenon can not be clearly observed if the size of the
plateau of the kovaton is less than the width of the compacton.
Note that the width of the signal during the collision is the sum of both the
width of the compacton and the kovaton before the interaction.
Also note the phase shift in the positions of both solitary waves after the
interaction.

\begin{figure}
 \begin{center}
  \includegraphics[width=\textwidth]{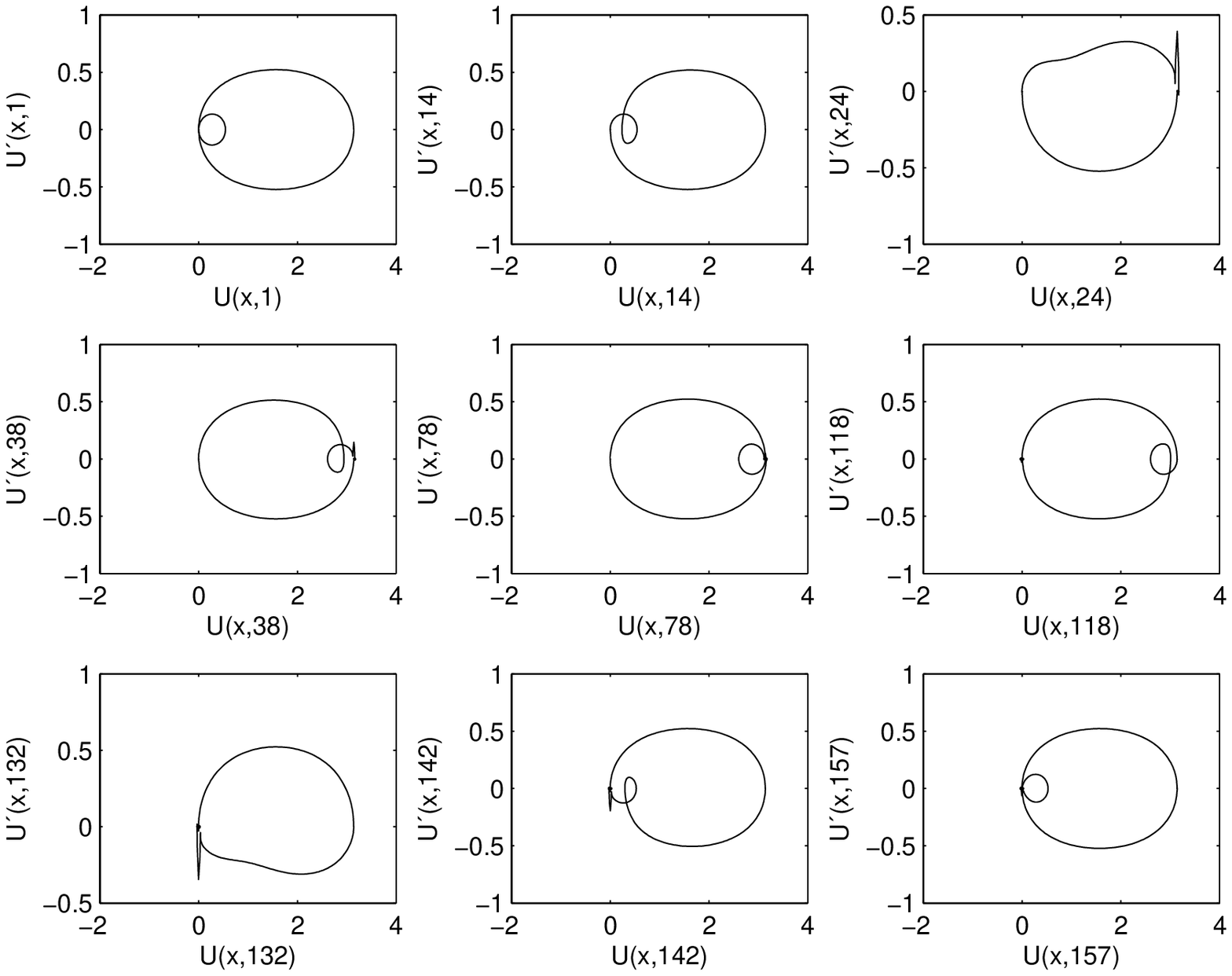}
 \end{center}
\caption{Sequence of the phase plane of a collision between a compacton
  with speed $c=0.2$ and a kovaton with plateau $m=40$,
  with $\mu=10^{-3}$, $c_0 = 2/\pi$, $x \in [0,200]$, $\Delta x = 0.05$,
  and $\Delta t = 0.01$.
 }
\label{fig:planofase}
\end{figure}

Compactons correspond to homoclinic orbits in the phase plane
$(u(\cdot,t),u_x(\cdot,t))$, at the critical point $(0,0)$, and kovatons to
heteroclinic orbits between the critical points $(0,0)$ and $(\pi,0)$; the
evolution in time of these orbits is a useful tool to study their interactions.
Figure~\ref{fig:planofase} shows the collision between a compacton with
speed $c=0.2$ and a kovaton with plateau length $m = 40$.
The interaction has actually three definite parts, the first being the entrance
of the compacton into the kovaton, the second being the travelling of the
compacton along the plateau of the kovaton, and the third being the exit
of the compacton from the kovaton.
In the top left plot of Fig.~\ref{fig:planofase} the phase plane shows two
ellipses, the largest one corresponding to the heteroclinic orbit of the
kovaton and the smallest one to the homoclinic orbit of the compacton, both
before the interaction begins.
Next, the compacton enters the rising kink of the kovaton, corresponding to
the left inner loop in the phase plane near the critical point $(0,0)$
(top center).
When the compacton reaches the plateau of the kovaton,
high slopes develop at the left side the plateau, causing the
appearance of a sharp peak in the phase plane near $(\pi,0)$ (top right).
Then the compacton is almost completely inside the plateau
of the kovaton, corresponding to the right inner loop in the phase plane near
$(\pi,0)$ (middle left).
When the entrance of the compacton has finished, the compacton travels reversed
all along the plateau of the kovaton, a remarkable situation that is
plotted by two orbits in the phase plane, the smallest one
corresponding now to an anticompacton on the plateau,
and the largest one to the kovaton (middle center).
When the compacton gets to the falling kink of the kovaton,
it recovers its original shape and the phase plane plots the right inner
loop at $(\pi,0)$ first (middle right),
the sharp peak at low amplitudes shortly afterwards (bottom left),
the left inner loop at $(0,0)$ (bottom center), and the final situation
with the two original orbits (bottom right).

\begin{figure}
 \begin{center}
  \includegraphics[width=12.5cm]{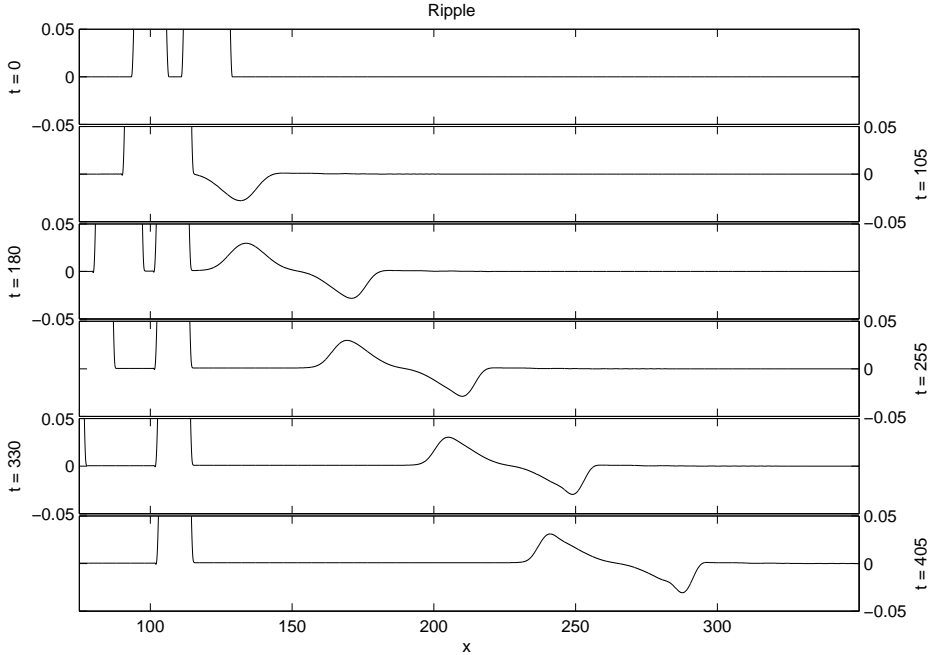}
\end{center}
  \caption{Sequence of zooms of the evolution of the ripple developed after a
   collision between two compactons, one with speed $c_1=0.5$
   (the left one in the first zoom) and the other one with speed
   $c_2=2/\pi$ (the right one in the same zoom), using
   $\mu=10^{-3}$, $c_0 = 0.5$, $\Delta x = 0.05$, and $\Delta t = 0.01$.}
  \label{fig:Residuo}
 \end{figure}

Another fact worth mentioning is the appearance of a zero-mass small amplitude
ripple after every collision either between two compactons or
between a compacton and a kovaton.
Figure~\ref{fig:Residuo} zooms it for two compactons with velocities
$c_1=0.5$ and $c_2=2/\pi$.
The first plot shows the situation previous to the collision, with the faster
(wider) compacton at the right.
In the next plot we can see the birth of the ripple,
only with its negative half part developed.
The following plots shows the evolution in time of this ripple, travelling to
the right in the computational frame at a speed equal to $c_0$, i.e.,
it remains at rest at the point where the collision begins.
This ripple is similar to those appearing in numerical simulations
of $K(2,2)$ compactons, except for the fact that there are no sharp fronts
inside (see~\cite{RusVillatoro2007b}).

\begin{figure}
 \begin{center}
  \includegraphics[width=12.5cm]{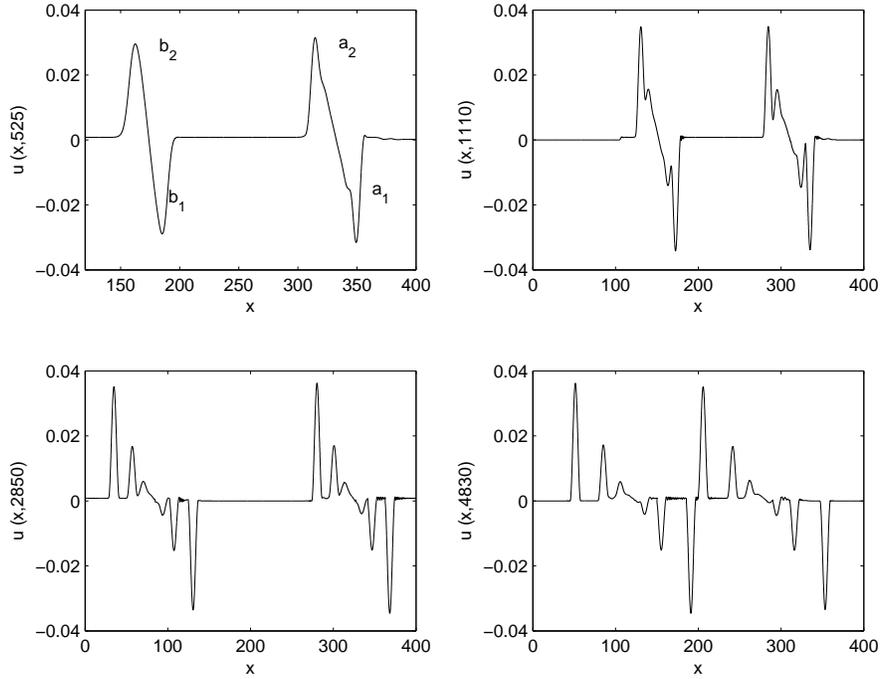}
\end{center}
  \caption{Evolution of the ripples generated after a collision between a
   compacton with speed $c=0.5$ and a kovaton with plateau length $m=40$, using
   $\mu=10^{-5}$, $\Delta x = 0.05$, $\Delta t = 0.01$, and $c_0=0$.
   The snapshots correspond to time $t=525$, 1110, 2850, and 4830.}
 \label{fig:ripples}
 \end{figure}

The residual appearing after a compacton-kovaton collision consists of
two separated ripples, as depicted in Fig.~\ref{fig:ripples} (top left plot),
the first one corresponding to the collision of the compacton with the rising
kink of the kovaton, labelled as $a_1$ and $a_2$ in the plot,
and the second one corresponding to the interaction of the
now anticompacton at the plateau of the kovaton with its falling kink,
labelled as $b_1$ and $b_2$.
Thus, $a_1$ ($a_2$) appears during the interaction between the negative
(positive) slope part of the compacton and the rising kink of the kovaton.
After the first ripple is generated,
the compacton travels reversed along the plateau, and then $b_1$ ($b_2$) appears
during the interaction between the positive (negative) slope part
of the now anticompacton and the falling kink of the kovaton.
It should be noted that the distance between the two ripples equals
the length of the plateau of the kovaton.
In the following plots of the same figure, as the evolution progresses,
the positive (negative) semi-ripples break down and decompose into trains of
positive (negative) pulses of decreasing amplitude, all of them with a width of
$4\, \pi$, corresponding to small amplitude compactons
(anticompactons).

\subsection{Compacton-anticompacton collisions}
\label{Results:Antiscollisions}

Collisions between compactons and anticompactons,
even adding the artificial viscosity term,
have proved to be unstable resulting in blow up, although
extensive numerical simulations have shown that the behaviour of the numerical
method for the propagation of anticompactons alone in terms of the error and the
invariants is the same as that of compactons with the same speed, discussed in
Section~\ref{ssec:OneCompacton}.

\begin{figure}
 \begin{center}
  \includegraphics[width=12.5cm]{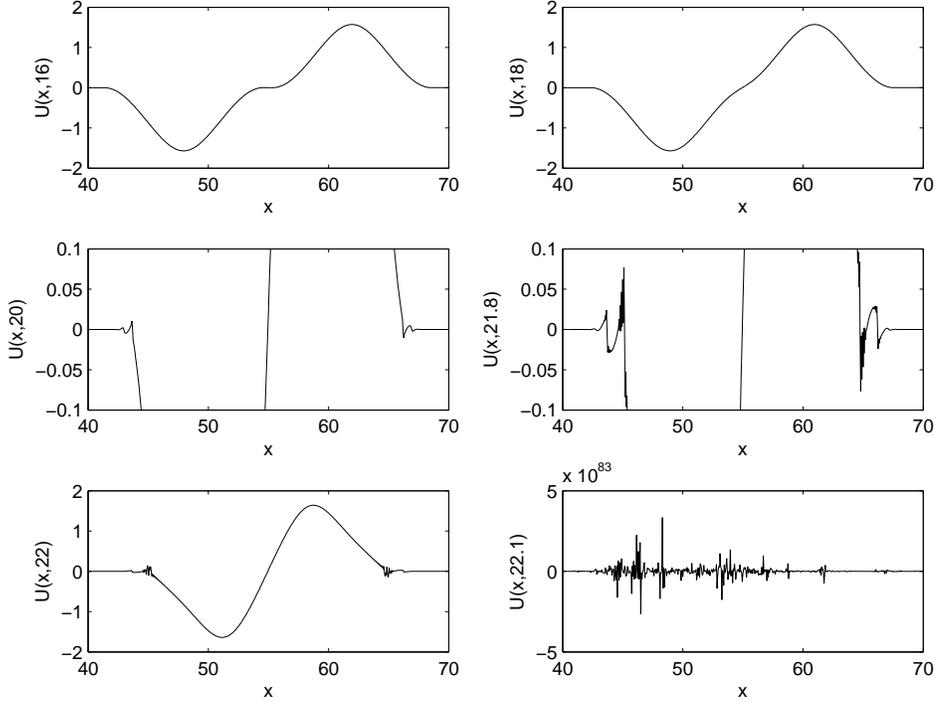}
\end{center}
  \caption{Snapshots of a compacton-anticompacton collision,
    both with $|c|=0.5$, at times $t=16$ (top left plot),
    $18$ (top right), $20$ (middle left), $21.8$ (middle right),
    $22$ (bottom left), and $22.1$ (bottom right), using
    $\Delta x= 0.05$, $\Delta t = 0.01$, $c_0=0$, and $\mu=10^{-4}$.}
  \label{fig:cp:acp}
 \end{figure}

Figure~\ref{fig:cp:acp} shows the collision between a
compacton with speed $c_1=0.5$ and an anticompacton with $c_2=-0.5$. The top plots are two snapshots previous
to the collision. In the middle ones, zoomed in, unexpected high frequency
phenomena that begin to occur at the external edges can be observed.
Conversely, the solution is apparently smooth at the central part.
The amplitude of these high frequency phenomena increases as the blow up
time approaches as shown in the bottom left zoomed out plot.
Finally, the last bottom right plot shows the blow up.

\begin{figure}
 \begin{center}
  \includegraphics[width=12.5cm]{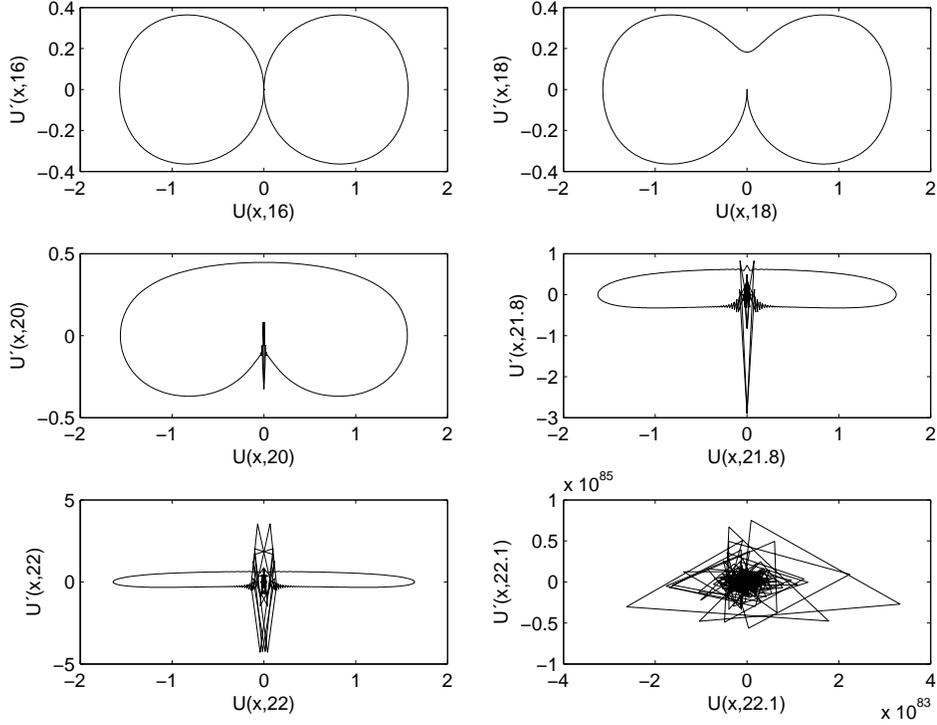}
\end{center}
  \caption{Sequence of phase plane plots of a compacton-anticompacton collision
   both with $|c|=0.5$ at times $t=16$ (top left plot),
   $18$ (top right), $20$ (middle left), $21.8$ (middle right),
   $22$ (bottom left), and $22.1$ (bottom right), using
   $\Delta x= 0.05$, $\Delta t = 0.01$, $c_0 = 0$, and $\mu=10^{-4}$.}
  \label{fig:cp:acp:pp}
 \end{figure}

Figure~\ref{fig:cp:acp:pp} shows the phase plane plots corresponding to
the same snapshots shown in Fig.~\ref{fig:cp:acp}.
The left homoclinic orbit (negative amplitudes) of the top left plot in
Fig.~\ref{fig:cp:acp:pp}
corresponds to the anticompacton, and the right one (positive amplitudes)
to the compacton.
The top right plot in Fig.~\ref{fig:cp:acp:pp}
show that as the travelling waves approach each other, the slope of the
solution increases at the
position where the compacton and anticompacton first met.
In the middle letf plot of Fig.~\ref{fig:cp:acp:pp}
a sharp peak in the bottom part of the melted
orbits can also be observed, corresponding to the high frequency phenomena
that begin to develop at the external edges of both travelling waves shown
in Fig.~\ref{fig:cp:acp} (middle left plot).
These high frequency phenomena increase at the external edges
causing the peak to grow in the phase plane as depicted in
in Fig.~\ref{fig:cp:acp:pp} (middle right and bottom left plots).
The final bottom right plot makes evident the blow up.

\begin{figure}
 \begin{center}
  \includegraphics[width=12.5cm]{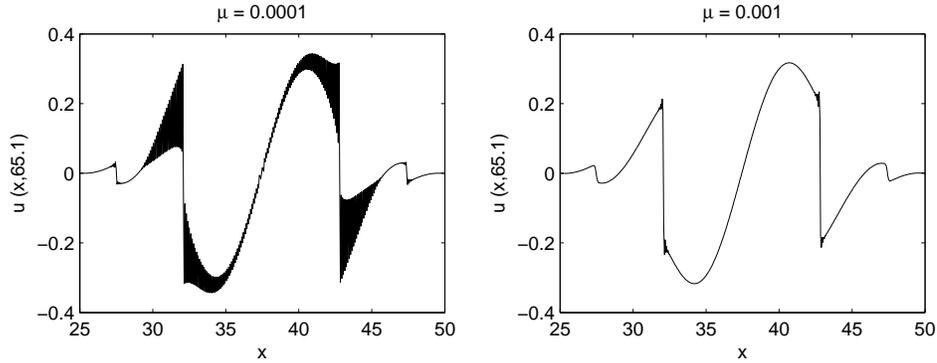}
\end{center}
  \caption{A compacton-anticompacton collision
   both with $|c|=0.5$ at time $t=65.1$ with $\mu=10^{-4}$ (left plot) and $10^{-3}$ (right one),
   using $\Delta x= 0.05$, $\Delta t = 0.01$, and $c_0 = 0$.}
  \label{fig:cp:acp:blowup}
 \end{figure}

The blow up during the compacton-anticompacton collision is due to the development of dispersive shock-like phenomena of increasing amplitude which appears in the regions of the solution with negative slope. Figure~\ref{fig:cp:acp:blowup} (left plot) shows the solution for $\mu=10^{-4}$ a few time steps before blow up; the use of a larger artificial viscosity (right plot) simply delays the blow up. The four shocks shown in Fig.~\ref{fig:cp:acp:blowup} present high frequency dispersion whose amplitude grows up as blow up time approaches. Our numerical simulations show that these phenomena is robust to changes in the parameters of the numerical method, $\Delta x$ and $\Delta t$, indicating its possible origin in the analytical behaviour of the solutions of the $K(\cos)$ equation.

\section{Conclusions}
\label{Conclusions}

The propagation of kovaton and compacton of the $K(\cos)$ equation has been
studied by means of a new Pad\'e numerical method. The good accuracy of the numerical method has been assessed by means of the error in the solution and the invariants for the propagation of compactons and kovatons. For compactons, small forward and backward wavepackets of radiation with a clear self-similar shape have been reported; the speed of the wave fronts of both radiations have been determined, being constant and independent of the grid spacing and time step of the numerical method. For kovatons, the main sources of the numerical error in the solution are the appearance of a small residual at the falling kink, whose area depends on the length of plateau, and the emission of a backward, square-like pulse whose length is also similar to that of the plateau.

The study of the collisions between compactly supported, solitary waves of the $K(\cos)$ equation requires the addition of small artificial viscosity for stability, as in the case of the $K(n,n)$ equation. For compactons, the collisions are very similar in both equations, showing a small amplitude residual decomposing in compacton-anticompacton pairs. The interaction between a compacton and a kovaton with a large enough plateau shows that the compacton passes through the rising kink, then it travels reversed like an anticompacton along the plateau, until reaching the falling kink of the kovaton, when it reemerges with its original shape. After the interaction two small residuals are generated, which decomposes in trains of compacton-anticompacton pairs. Finally, the collisions between anticompactons and compactons result in the blow up of the solution due to the grow up in amplitude of dispersive shock-like structures in the solution.

Present results show the great interest of the numerical study of the interactions of compactly supported solitary waves of generalizations of the $K(n,n)$ equation, as those presented in Ref.~\cite{RusVillatoro2009b}, since only a few have been numerically studied in detail. Furthermore, the use of other numerical methods, like finite volume methods for nonconvex conservation laws, is promising since the presence of shock-like phenomena can be considered a common feature of all the evolution equations with nonlinear dispersion.

\section*{Acknowledgments}

The research reported here was partially supported by Projects
MTM2010--19969 and TIN2008--05941 of the Ministerio de Ciencia e
Innovaci\'on of Spain, and by Project TIC-6083 of the Consejer\'ia de Econom\'ia, Innovaci\'on y Ciencia of the Junta de Andaluc\'ia.


\newcommand{\bookrefchap}[6]{#1, ``{\em {#2}}," #3, Chap. #4, pp. #5 (#6).}
\newcommand{\bookref}[4]{#1, ``{\em {#2}}," (#4) #3.}

\newcommand{\paperref}[8]{#1, #2, {#8} #4 (#7) #6.}
\newcommand{\paperrefart}[9]{#1, #2, {#9}, #4, (#8) Art. #6.}

\newcommand{\paperrep}[5]{#1, ``#2", {\em {#3}}, {#4} (#5).}

\newcommand{\procref}[7]{#1: ``#2" in ``{\em #3}" edited by #4, #5, pp. #6,
#7.}

\newcommand{\procrefvol}[9]{#1: ``#2" in ``{\em #3}", #4, edited by #5, #6, vol.
#7, pp. #8, #9.}

\end{document}